\documentclass{Interspeech}

\usepackage{amsmath,amssymb,amsthm}
\usepackage{amsfonts} % mathbb
\usepackage{gensymb}  % for \degree
\usepackage{multicol}
\usepackage{multirow}
\usepackage{pifont}% http://ctan.org/pkg/pifont

\usepackage{xpatch,xcolor}
\usepackage{hyperref}

\usepackage{bbm}
\usepackage{booktabs}
\usepackage{makecell}
\interspeechcameraready

\title{TTMBA: Towards Text To Multiple Sources Binaural Audio Generation}

\author[affiliation={1}]{Yuxuan}{He}
\author[affiliation={1}]{Xiaoran}{Yang}
\author[affiliation={2}]{Ningning}{Pan}
\author[affiliation={1}]{Gongping}{Huang}

\affiliation{School of Electronic Information}{Wuhan University, Wuhan, Hubei}{China}
\affiliation{School of Computing and Artificial Intelligence}{Southwestern University of Finance and Economics, Chengdu, Sichuan}{China}
\email{yuxuanhe@whu.edu.cn, yxr888@whu.edu.cn, nnpan@swufe.edu.cn, gongpinghuang@whu.edu.cn
\thanks{This work was supported in part by the National Natural Science Foundation of China (NSFC) under Grant 62471340 and in part by the Guangdong Basic and Applied Basic Research Foundation under Grant 2025A1515010226.}}

\keywords{Text to binaural audio generation, temporal control, spatial control, binaural audio rendering, multisource}

\usepackage{comment}

\begin{document}

\maketitle

\begin{abstract}
Most existing text-to-audio (TTA) generation methods produce mono outputs, neglecting essential spatial information for immersive auditory experiences. To address this issue, we propose a cascaded method for text-to-multisource binaural audio generation (TTMBA) with both temporal and spatial control. First, a pretrained large language model (LLM) segments the text into a structured format with time and spatial details for each sound event. Next, a pretrained mono audio generation network creates multiple mono audios with varying durations for each event. These mono audios are transformed into binaural audios using a binaural rendering neural network based on spatial data from the LLM. Finally, the binaural audios are arranged by their start times, resulting in multisource binaural audio. Experimental results demonstrate the superiority of the proposed method in terms of both audio generation quality and spatial perceptual accuracy.
\end{abstract}

\section{Introduction}
With the rise of rapid technological innovation, the demand for high-fidelity binaural audio generation has surged in fields such as virtual reality (VR), augmented reality (AR), and mixed-reality systems, particularly for education and entertainment. While deep generative models have advanced audio generation, previous research has primarily focused on mono audio, limiting their effectiveness in scenarios requiring spatial acoustic signatures.
In recent years, extensive research has been conducted, and significant progress has been achieved in this field. For instance, Diffsound~\cite{10112585}, the first diffusion-based audio generation model, uses a pre-trained VQ-VAE~\cite{van2017Neural} to map mel-spectrograms into discrete tokens, which are generated with a diffusion model. AudioGen~\cite{DBLP:conf/iclr/KreukSPSDCPTA23} adopts a similar VQ-VAE with an autoregressive model in a discrete waveform space. In contrast, AudioLDM~\cite{pmlr-v202-liu23f, audioldm2-2024taslp} employs a continuous latent diffusion model (LDM)~\cite{Rombach_2022_CVPR}, yielding better audio quality. While these models effectively control audio duration, they face challenges with multisource audio control, particularly in scenarios requiring precise spatial accuracy

To address these challenges, recent studies have focused on improving temporal and spatial control in audio generation. Make-An-Audio 2~\cite{DBLP:journals/corr/abs-2305-18474}
         and TangoFlux~\cite{DBLP:journals/corr/abs-2412-21037} prioritize temporal information for variable-length audio generation. Make-An-Audio 2 uses an LLM and temporal encoder for coherent audio generation but incurs high computational costs due to LDM. TangoFlux, a lightweight model for duration control, generates binaural audio by duplicating single-channel recordings, resulting in nearly identical outputs and lacking spatial control through text. To enhance spatial accuracy, we first utilize TangoFlux and subsequently apply binaural rendering with directional cues.
BEWO~\cite{DBLP:journals/corr/abs-2410-10676} is a latent diffusion model that utilizes text embedding and azimuth information to generate spatial audio. Similarly, Immersive Diffusion~\cite{DBLP:journals/corr/abs-2410-14945} combines ELSA~\cite{DBLP:conf/nips/DevnaniSATMTSS24} with the Diffusion Transformer to produce spatial audio. Both models accept text inputs and generate spatial audio, but they lack precise temporal control and neglect listener anatomy, both of which are essential for authentic binaural perception.

\begin{figure}[t!]
	\centering
	\includegraphics[width=\linewidth]{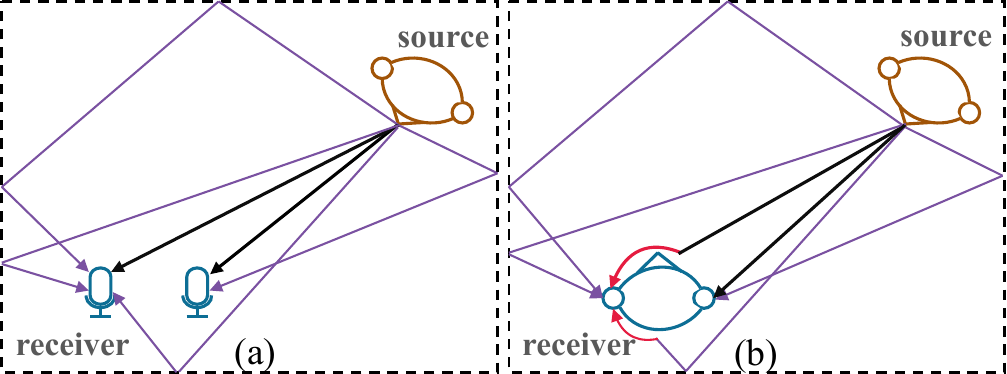}
    \vspace{-12pt}
        \caption{Illustration of acoustic propagation models: (a) Conventional geometric acoustic simulation based solely on room impulse responses; (b) Binaural modeling incorporating listener-specific cues.}
	\label{fig1}
        \vspace{-16pt}
\end{figure}

% \begin{figure*}[t!]
% 	\centering
% 	\includegraphics[width=\linewidth]{figs/fig2.pdf}
%         \caption{The framework of the proposed text-to-multisource binaural audio generation network.}
% 	\label{fig2}
% \end{figure*}

Another class of methods that has garnered significant attention involves rendering mono audio to binaural audio. Traditional simulators, such as Pyroomacoustics~\cite{8461310} efficiently model acoustic propagation between sources and receivers, as shown in Fig.~\ref{fig1}~(a). However, these methods mainly focus on room impulse responses (RIRs) and neglect crucial factors such as the listener’s pinnae, head, and torso, all of which are essential for authentic binaural perception~\cite{Wright1974Pinna}. To address this limitation, Head Related Transfer Functions (HRTF) and Binaural Room Transfer Functions (BRTF) model these cues ~\cite{Wightman1989Headphone}. The NIIRF framework~\cite{10448477} approximates direction-dependent HRTF coefficients with cascaded IIR filters but suffers from spectral errors. Neural networks like BinauralGrad~\cite{DBLP:conf/nips/LengCGLCTMHLQzL22} and NFS~\cite{10095685} generate binaural audio, with NFS excelling in both static and dynamic scenarios in the Fourier space. These methods account for the listener’s anatomy, as shown in Fig.~\ref{fig1}~(b).

\begin{figure*}[t!]
	\centering
	\includegraphics[width=\linewidth]{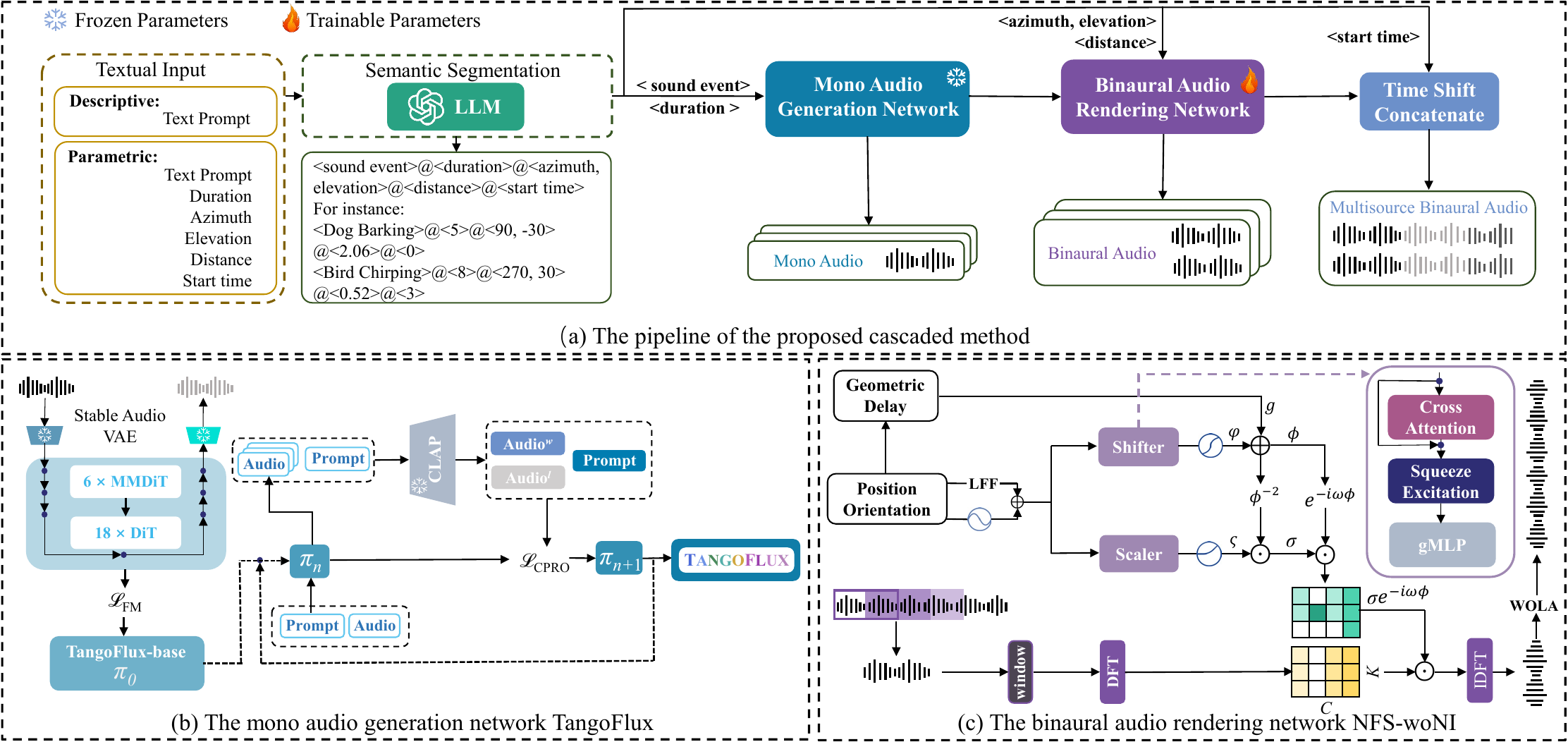}
    \vspace{-12pt}
        \caption{The framework of the proposed text-to-multisource binaural audio generation network.}
	\label{fig2}
    \vspace{-16pt}
\end{figure*}

In this paper, we propose a text-to-multisource binaural audio generation model (TTMBA)\footnote{Demo page:\url{https://25219.github.io/1.github.io/}} that synthesizes binaural audio with controllable spatial (azimuth, elevation, distance) and temporal (duration, start time) attributes. 
The model first employs a pre-trained large language model (LLM) to extract source information, followed by an audio generation network that synthesizes individual mono streams, and a binaural rendering network simulates directional effects. 
The contributions of this work include: \begin{itemize} 
\item The first text-conditioned binaural audio generation model with control over duration, start time, and location. 
\item The use of a LLM to extract source location from text, relying on physical principles when cues are absent. 
\item The developed method achieves strong performance with low computational cost in both quantitative and perceptual evaluations.
\end{itemize}

\section{Method}

This work proposes a cascaded pipeline for generating multisource, duration-controllable binaural audio. It uses a large language model (LLM), GPT-$4$o\footnote{\url{https://openai.com/blog/chatgpt}}, to segment descriptive or parametric text descriptions into multiple sets of structured datas in a standard format of \textless sound event\textgreater @\textless duration\textgreater @\textless azimuth, elevation\textgreater @\textless distance\textgreater @\textless start time\textgreater, inferring spatial details through commonsense reasoning when position description is missing (e.g., Dog barking typically comes from the downside of human). Sound events and durations are used in TangoFlux\footnote{\url{https://github.com/declare-lab/TangoFlux}} to generate mono audios, while azimuth, elevation, and distance information guide binaural rendering to produce spatially accurate binaural audio. Finally, these binaural audios are rearranged into one output based on the start time. The pipeline is shown in Fig.~\ref{fig2}~(a). 

\subsection{Text to Mono Audio Generation}

We adopt TangoFlux for mono audio generation. As shown in Fig.~\ref{fig2}~(b),  the text-to-mono audio generation network incorporates a unique combination of $6$ Multimodal Diffusion Transformer (MMDiT)~\cite{DBLP:conf/icml/EsserKBEMSLLSBP24} and $18$ Diffusion Transformer (DiT)~\cite{DBLP:conf/iccv/PeeblesX23} blocks, which is conditioned on both text embedding and duration embedding to generate high-quality audio. Initially, a variational autoencoder (VAE) is employed to encode the latent representations of audio, guiding the network to learn a rectified flow trajectory. Throughout the training process, the VAE is kept frozen.

Given a latent representation $\mathbf{x}_1$ of an audio sample encoded by the VAE and a noise input $\mathbf{x}_0 \sim \mathcal{N}(\mathbf{0}, \mathbf{I})$,
we construct an intermediate state $\mathbf{x}_t$ for $t \in [0,1]$, where $t$ denotes the flow matching timestep. The network learns to predict a velocity, $\mathbf{v}_t = \frac{d\mathbf{x}_t}{dt}$, guiding $\mathbf{x}_t$ toward the target $\mathbf{x}_1$. Rectified flows (RFs)~\cite{DBLP:conf/iclr/LiuG023} are then applied to transform the noise distribution into the target distribution, following straight-line trajectories.
The model $\mathbf{u}$, where the parameters are denoted as $\theta$, is trained to directly regress the predicted velocity $\mathbf{u}(\mathbf{x}_t, t; \theta)$ toward the ground-truth velocity $\mathbf{v}_t$, according to the loss function:
\begin{equation}
\mathcal{L}_{\mathrm{FM}} = \mathbb{E}_{\mathbf{x}_1,\mathbf{x}_0,t}\Vert \mathbf{u}(\mathbf{x}_t,t;\theta)-\mathbf{v}_t \Vert^2.
\label{eq:flow_loss}
\end{equation}

The network is first pretrained as the initial base model $\pi_{0}$. Sebsequently, CLAP-Ranked Preference Optimization (CRPO) is leveraged to refine the model iteratively, which aligns checkpoint $\pi_{n}$ into checkpoint $\pi_{n+1}$, starting from $n = 0$. Here, the CLAP model functions as a proxy reward model to assess audios according to how well they match the text input. The loss function for preference optimization, $\mathcal{L}_{\text{DPO-FM}}$, is established to optimize the alignment between the generated audio and the given text description, is defined as: 
\begin{equation}
\begin{aligned}
\mathcal{L}_{\text{DPO-FM}} &= -\mathbb{E}_{t \sim \mathcal{U}(0,1), \mathbf{x}^w, \mathbf{x}^l} \log \sigma \left( -\beta \left( \Vert \mathbf{u}(\mathbf{x}_t^w, t; \theta) - \mathbf{v}_t^w \Vert_2^2 \right.\right. \\
&\quad \left. \left. - \Vert \mathbf{u}(\mathbf{x}_t^l, t; \theta) - \mathbf{v}_t^l \Vert_2^2 - \left( \Vert \mathbf{u}(\mathbf{x}_t^w, t; \theta_{\text{ref}}) - \mathbf{v}_t^w \Vert_2^2 \right.\right.\right. \\
&\quad \left. \left. \left. - \Vert \mathbf{u}(\mathbf{x}_t^l, t; \theta_{\text{ref}}) - \mathbf{v}_t^l \Vert_2^2 \right) \right) \right). \label{eq:dpo_rearrange}
\end{aligned}
\end{equation}
where $ \mathbf{x}^w_t $ and $ \mathbf{x}^l_t $ correspond to the winning and losing audios evaluated by CLAP, respectively, and $\theta_{\text{ref}}$ denotes the parameters of the current base model.

To mitigate the risk of overoptimization and maintain stable training, the total loss function $\mathcal{L}_{\text{CRPO}}$ integrates the preference optimization loss $\mathcal{L}_{\text{DPO-FM}}$ in (\ref{eq:dpo_rearrange}) with the flow matching loss $\mathcal{L}_{\text{FM}}$ in (\ref{eq:flow_loss}). This combination is formulated as:

\begin{equation}
\mathcal{L}_{\text{CRPO}} := \mathcal{L}_{\text{DPO-FM}} + \mathcal{L}_{\text{FM}}.
\end{equation}

This integrated loss function preserves the semantic and structural integrity of the generated audio while refining preference rankings, resulting in a balanced and resilient optimization. With iterative alignment, a hybrid architecture, and rectified flow, the network generates high-quality, tailored audio that aligns accurately with textual prompts and specified lengths.

\subsection{Mono to Binaural Audio Rendering}

As illustrated in Fig.~\ref{fig2}~(c), the Binaural Audio Rendering Network converts mono audio into binaural audio using Fourier transform theories. When sound propagates, delays and energy changes occur due to attenuation and absorption, which correspond to phase shifts and magnitude reductions in the Fourier domain. The Discrete Fourier Transform (DFT) is applied to the mono signal to compute its frequency-domain representation, where $\omega_k = \frac{2\pi k}{K}$, with $k \in \{ 0,1,\ldots,K-1 \}$ and $\textbf{X} \in \mathbb{C}^K$. The Fourier shift theorem relates temporal shifts to linear phase shifts, enabling the network to render sound propagation effects like delays and energy alterations accurately.

Building on this, the network aims to predict the magnitude reductions $\sigma \in \mathbb{R}^{C \times K}$ and phase shifts $\phi \in \mathbb{R}^{C \times K}$, where $C$ represents the number of channels for the left (L) and right (R) ears, respectively. Subsequently, the network output, i.e., $\hat{\textbf{X}}_\text{L}(k)$ and $\hat{\textbf{X}}_\text{R}(k)$, are obtained by applying these predicted magnitude scales and phase shifts to the mono spectrum $\mathbf{X}(k)$:
\begin{align}
    \hat{\mathbf{X}}_{\text{L}}(k) &= \sum_{c=0}^{C-1} \sigma_{\text{L}}(c,k) e^{-i \omega_k \phi_{\text{L}}(c,k)} \mathbf{X}(k), \notag \\
    \hat{\mathbf{X}}_{\text{R}}(k) &= \sum_{c=0}^{C-1} \sigma_{\text{R}}(c,k) e^{-i \omega_k \phi_{\text{R}}(c,k)} \mathbf{X}(k). \label{eqn:spec}
\end{align}

The network takes three inputs: mono audio, the source’s position with orientation, and frame-wise geometric delay. The position \textless azimuth, elevation\textgreater and \textless distance\textgreater~are transformed into a spatial position $p \in \mathbb{R}^3$ and orientation $q \in \mathbb{R}^4$, while geometric delay $g$ is calculated from direct-path distances, as shown in Fig.~\ref{fig1}. The mono audio is segmented and windowed with a Hanning window. Position and orientation are encoded using sinusoidal encoding and learned Fourier features ({\tt\footnotesize LFF}). Two parallel modules, {\tt\footnotesize Scaler} and {\tt\footnotesize Shifter}, predict spectral magnitude coefficients and phase offsets. Both modules use cross-attention~\cite{vaswani2017attention}, squeeze-and-excitation~\cite{hu2018squeeze}, and gMLP~\cite{liu2021pay} to encode spatial and orientational information. The outputs pass through nonlinear activation functions, $\varsigma$ (strictly positive) and $\varphi$ (capped under half the frame length).

The geometric delay $g$ computed from the warp field is added to $\varphi$ for each frame, yielding $\phi=\varphi+g$. To ensure that energy is inversely proportional to the square of the distance, we normalize $\varsigma$ by dividing it element-wise by $\phi^{2}$, resulting in the scale parameter $\sigma=\varsigma\odot\phi^{-2}$. We then compute the frame-wise multichannel spectrum $\sigma e^{(-i\omega\phi)}$, which is multiplied with the source in the Fourier domain, as shown in \eqref{eqn:spec}. Next, IDFT is applied to $\hat{\mathbf{X}}\in\mathbb{C}^{C\times K}$, and the channels are linearly combined into a single stream. The final output $\hat{\mathbf{x}}$ is generated using a weighted overlap-add (WOLA) synthesis method, without ambient noise injection to enhance quality and mimic real-world binaural recordings, which differentiates it from the initial NFS. This network is referred to as NFS-woNI.

To ensure the generated binaural audio closely matches the target signal $\mathbf{y}$, the loss function of the network is defined as:
\begin{equation}
\begin{aligned}
\mathcal{L}(\hat{\textbf{x}}, \textbf{y}) &= \lambda_1 \Vert\hat{\textbf{x}} - \textbf{y}\Vert_2  + \lambda_2 \Vert\angle \hat{\textbf{X}} - \angle \textbf{\textbf{Y}}\Vert_1 \\
& + \lambda_3 \Vert \text{IID}(\hat{\textbf{x}}) - \text{IID}(\textbf{y})\Vert_2  + \lambda_4 \text{MRSTFT}(\hat{\textbf{x}}, \textbf{y}),
\end{aligned}
\label{eq:nfs_loss}
\end{equation}
where $\text{IID}$ quantifies the interaural intensity difference, and $\text{MRSTFT}$~\cite{9053795} represents the multi-resolution short-time Fourier transform loss. The four components in \eqref{eq:nfs_loss} are denoted as $\ell_2$, $\mathcal{L}_{\text{phs}}$, $\mathcal{L}_{\text{IID}}$, and $\mathcal{L}_{\text{STFT}}$, respectively.

\begin{table*}[t!]
    \caption{Comparison of different TTA models and the proposed method evaluated by $1$ channel across various metrics. The highest score for each metric is marked with an asterisk ($\cdot^*$) and the second highest score is marked with a dagger ($\cdot^\dagger$).}
    \label{tab:comparison_1}
    \centering
    \renewcommand{\arraystretch}{0.9}
    \begin{small}
    \begin{tabular*}{\linewidth}{@{\extracolsep{\fill}} l c c c c c c c }
        \toprule
        \small Method & \small FD $\downarrow$ & \small KL $\downarrow$ & \small IS $\uparrow$ & \small CLAP $\uparrow$ & \small MOS-Q $\uparrow$ & \small MOS-F $\uparrow$ & \small Inference Time (s) $\downarrow$ \\
        \midrule
        \small AudioLDM & 21.06 & 1.43 & 9.89 & 0.54 & 3.07 & 2.61 & 41.55 \\
        \small Make-An-Audio 2 & \textbf{13.43}$^*$ & 1.38 & 10.58 & 0.55 & 3.36 & 3.32 & 27.77 \\
        \small TangoFlux & \textbf{17.74}$^\dagger$ & \textbf{1.33}$^*$ & \textbf{13.94}$^*$ & \textbf{0.61}$^*$ & \textbf{4.64}$^*$ & \textbf{4.79}$^*$ & \textbf{2.22}$^*$ \\
        \small \textbf{TangoFlux-NFS-woNI} & 17.81 & \textbf{1.35}$^\dagger$& \textbf{11.79}$^\dagger$ & \textbf{0.58}$^\dagger$ & \textbf{4.57}$^\dagger$ & \textbf{4.57}$^\dagger$ & \textbf{2.58}$^\dagger$ \\
        \bottomrule
    \end{tabular*}
    \end{small}
\end{table*}
\begin{table*}[t!]
    \caption{Comparison of different binaural audio rendering methods evaluated by $2$ channels across various metrics. The highest score for each metric is marked with an asterisk ($\cdot^*$) and the second highest score is marked with a dagger ($\cdot^\dagger$).}
    \label{tab:comparison_2}
    \centering
    \renewcommand{\arraystretch}{0.9}
    \begin{small}
    \begin{tabular*}{\linewidth}{@{\extracolsep{\fill}} l c c c c c c c }
        \toprule
        \small Method & \small$\ell_2$ $\downarrow$ & \small $\mathcal{L}_{\text{mag}}$ $\downarrow$ & \small $\mathcal{L}_{\text{phs}}$ $\downarrow$ & \small$\mathcal{L}_{\text{STFT}}$ $\downarrow$ & \small PESQ $\uparrow$ & \small MOS-Q $\uparrow$ & \small MOS-P $\uparrow$ \\
        \midrule
        \small TangoFlux-Pyroom & 0.10 & 0.92 & \textbf{1.57}$^\dagger$ & 2.05 & \textbf{2.58}$^*$ & \textbf{4.23}$^\dagger$ & 3.48 \\
        \small TangoFlux-NIIRF & 0.12 & 0.72 & \textbf{1.57}$^*$ & \textbf{1.89}$^\dagger$ & 2.25 & 4.00 & 2.86 \\
        \small TangoFlux-BinauralGrad & \textbf{0.06}$^*$ & \textbf{0.67}$^\dagger$ & 1.57 & 3.37 & 1.77 & 3.19 & \textbf{3.95}$^\dagger$ \\
        \small \textbf{TangoFlux-NFS-woNI} & \textbf{0.08}$^\dagger$ & \textbf{0.57}$^*$ & 1.58 & \textbf{1.61}$^*$ & \textbf{2.43}$^\dagger$ & \textbf{4.71}$^*$ & \textbf{4.57}$^*$ \\
        \bottomrule
    \end{tabular*}
    \end{small}
    \vspace{-12pt}
\end{table*}
\section{Experimental Setup}

\subsection{Datasets and Baselines}

To evaluate the quality of the generated mono audios, we conduct experiments using the test set of the AudioCaps dataset~\cite{DBLP:conf/naacl/KimKLK19}, which consists of 4,368 audio-text pairs. Each audio clip is 10 seconds long with a 16 kHz sampling rate. For baseline comparisons, we choose AudioLDM and Make-An-Audio 2. Specifically, we use the improved version of AudioLDM, AudioLDM\_16k\_crossattn\_t5, as our selected checkpoint.
To ensure fair comparison of duration-controllable audio generation, all audio clips are uniformly truncated or padded to 8 seconds, preserving as much audio information as possible.

As for the binaural audio generation, the binaural rendering network is trained utilizing the Binaural Speech Benchmark Dataset~\cite{richard2020neural}, which contains approximately 2 hours of mono-binaural paired audio data recorded on a KEMAR dummy head. 
For testing, we employ the CIPIC dataset~\cite{969552}, selecting two sets of HRTFs also recorded on a KEMAR dummy head to convolve with the generated mono audio, creating reference audios for the next stage. These reference audios serve as the benchmark for calculating both the objective and subjective metrics of our proposed method and the baselines. We compare the NFS-woNI with Pyroomacoustics, NIIRF, and BinauralGrad, each of which takes the generated mono audio as input. These methods are referred to as TangoFlux-NFS-woNI, TangoFlux-Pyroom, TangoFlux-NIIRF, and TangoFlux-BinauralGrad, respectively.

\subsection{Training Details}

We use the pre-trained TangoFlux model to generate mono audios. When rendering binaural audios, both the channel count $C$ and the dimensions for sinusoidal encoding are set to 128. The model is refined using the RAdam optimizer~\cite{DBLP:conf/iclr/LiuJHCLG020} over 16 training epoches, employing a batch size of 6. The initial learning rate is set to \(10^{-3}\) and is decayed by a factor of 0.9 at the end of each epoch. The parameters in \eqref{eq:nfs_loss} are as follows: $\lambda_1 = 10^{3}$, $\lambda_3 = 10^{1}$, and $\lambda_2 = \lambda_4 = 1$. Notably, the network is trained without noise injection to preserve audio quality—a slight modification compared to the NFS. 

\section{Results}

\subsection{Mono Audio Generation Evaluation}

We assess the generated audio using both objective and subjective metrics. Objective metrics include Fréchet Distance (FD) ~\cite{8682475}, Kullback-Leibler divergence (KL)~\cite{Kullback1951OnIA}, Inception Score (IS)~\cite{salimans2016improvedtechniquestraininggans}, and CLAP score~\cite{10095889}. FD measures similarity to target audio, KL quantifies differences in acoustic event posteriors, IS evaluates diversity and specificity, and CLAP score calculates cosine similarity between audio embeddings and their textual descriptions. Lower FD and KL values indicate better alignment, while higher IS and CLAP scores reflect greater diversity and stronger text-audio alignment. For subjective evaluation, participants with normal hearing rate the mono audios using Mean Opinion Score (MOS) for audio quality (MOS-Q) and text-audio alignment (MOS-F).

We first compare the performance of TangoFlux in the first stage with two other TTA models, namely AudioLDM and Make-An-Audio 2. To verify that the processing in the second stage does not compromise audio quality, we also evaluate metrics of the binaural audio output. Specifically, one channel is extracted from the binaural audio for metric assessment.
Table~\ref{tab:comparison_1} demonstrates that TangoFlux outperforms other TTA models across all metrics, with a particularly significant reduction in the average inference time per audio sample, among which all inference times are computed on the NVIDIA GeForce RTX 4090 GPU. Additionally, the result indicate that the audio generated by the proposed TangoFlux-NFS-woNI exhibits similar performance to TangoFlux, with minimal distortion in audio quality due to binaural audio rendering. The MOS-Q and MOS-F scores confirm that the audio generated by the proposed method maintains high quality and accurate text-audio alignment.

\subsection{Binaural Audio Rendering Evaluation}
In our evaluation of binaural audio, we employ both objective and subjective metrics to comprehensively assess the quality and spatial accuracy of the generated audio. For the objective evaluation, we utilize five metrics: $\ell_2$, $\mathcal{L}_{\text{mag}}$, $\mathcal{L}_{\text{phs}}$, $\mathcal{L}_{\text{STFT}}$, and (PESQ)~\cite{941023}, where $\mathcal{L}_{\text{mag}}$ quantifies the spectral magnitude error between the generated and reference audio, while $\ell_2$, $\mathcal{L}_{\text{phs}}$, and $\mathcal{L}_{\text{STFT}}$ are defined in \eqref{eq:nfs_loss}. PESQ measures the perceptual quality of the audio. Subjectively, we use MOS-Q and MOS-P to assess overall audio quality and spatial accuracy, respectively. For MOS-P, listeners compare the reference audio with four generated audios, scoring each based on the perceived similarity of the sound source location to the reference.

\begin{figure}[h!]
	\centering
	\includegraphics[width=\linewidth]{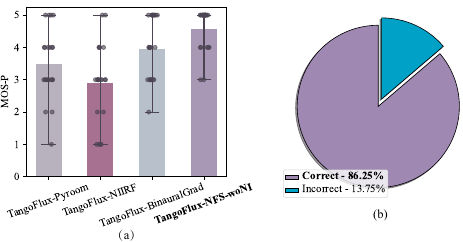}
    \vspace{-12pt}
        \caption{Subjective evaluations of the generated binaural audio: (a) The comparison of MOS-P distribution and average scores across the four methods; (b) Percentage of correct answers in the direction perception test.}
	\label{fig3}
    \vspace{-16pt}
\end{figure}

We then evaluate the performance of different binaural audio rendering methods in the second stage.  As shown in Table~\ref{tab:comparison_2}, the proposed TangoFlux-NFS-woNI obtains the best $\mathcal{L}_{\text{mag}}$ and $\mathcal{L}_{\text{STFT}}$ scores and the second-best $\ell_2$ and PESQ scores accross objective metrics. It also outperforms all other methods in subjective metrics, MOS-Q and MOS-P, which confirm that the binaural audios generated by TangoFlux-NFS-woNI maintain high quality and accurately preserves source positioning. In contrast, TangoFlux-BinauralGrad demonstrates good performance in MOS-P but get low score in MOS-Q, indicating the model is not robust in static sound source. TangoFlux-Pyroom performs well in MOS-Q but shows not well in MOS-P, indicating binaural audio requires further consideration of spectral distortions resulting from interactions with the listener’s pinnae, head, and torso. The distribution of MOS-P collected from the audience is shown in the Fig.~\ref{fig3}~(a), where the height of each bar represents the average score of the corresponding method, and each scatter point represents an individual rating.

% \begin{figure}[h!]
% 	\centering
% 	\includegraphics[width=\linewidth]{figs/fig3.pdf}
%         \caption{Subjective evaluations of the generated binaural audio: (a) The comparison of MOS-P distribution and average scores across the four methods; (b) Percentage of correct answers in the direction perception test.}
% 	\label{fig3}
%     \vspace{-12pt}
% \end{figure}

We also evaluate the accuracy of the sound source directions perceived by listeners based on the binaural outputs of TangoFlux-NFS-woNI. Specifically, listeners are asked to identify the direction of two sound sources by selecting: 1) left or right, 2) front or rear, and 3) above or below (e.g., dog barking is left, rear, below; bird chirping is right, front, above).  As shown in Fig.~\ref{fig3}~(b), the results demonstrate that most participants accurately identified the sound source directions, achieving an 86.25\% accuracy rate. This confirms the capability of the proposed method to render sound sources in the desired spatial directions.

\section{Conclusion}

We propose a cascaded lightweight neural network for text-to-multisource binaural duration controllable audio generation. The model uses GPT-$4$o to extract sound events and sources from text input, TangoFlux to generate mono audios, a Fourier transform network to render high quality binaural audios. Experimental results confirm that the proposed method generate high quality binaural audio with both temporal and spatial control, specifically, audio quality perform best and over 85\% accuracy in source locations according to user-defined conditions. %We expect improving the proposed method using anthropology parameters to generate personalized binaural audio.

\newpage
\bibliographystyle{IEEEtran}
\bibliography{mybib}

\end{document}